\documentclass[10pt,aps,pra,twocolumn,showpacs,superscriptaddress,showpacs,tightenlines,longbibliography]{revtex4-2}

\usepackage{amsthm}
\usepackage{amsmath}            
\usepackage{amssymb}            
\usepackage{mathtools}
\usepackage{graphicx}           
\usepackage{multirow}

\graphicspath{ {./images/} }

\usepackage[colorlinks,citecolor=blue,urlcolor=blue,bookmarks=false,hypertexnames=true]{hyperref}

\newcommand{\ket}[1]{| #1 \rangle}

\usepackage[normalem]{ulem}

\newcommand{\comt}[1]{}

\newcommand{\avg}{\textrm{avg}}

\begin{document}
\title{
Experimental simulation of postselected closed timelike curves for\\
decoding scrambled quantum information
}

\author{Yi-Te Huang}
\affiliation{Department of Physics, National Cheng Kung University, Tainan 701401, Taiwan}
\affiliation{Center for Quantum Frontiers of Research and Technology (QFort), Tainan 701401, Taiwan}
\affiliation{RIKEN Center for Quantum Computing, RIKEN, Wakoshi, Saitama 351-0198, Japan}

\author{Hsiang-Wei Huang}
\affiliation{Department of Physics, National Cheng Kung University, Tainan 701401, Taiwan}
\affiliation{Center for Quantum Frontiers of Research and Technology (QFort), Tainan 701401, Taiwan}

\author{Jhen-Dong Lin}
\affiliation{Department of Physics, National Cheng Kung University, Tainan 701401, Taiwan}
\affiliation{Center for Quantum Frontiers of Research and Technology (QFort), Tainan 701401, Taiwan}

\author{Adam Miranowicz}
\affiliation{RIKEN Center for Quantum Computing, RIKEN, Wakoshi, Saitama 351-0198, Japan}
\affiliation{Institute of Spintronics and Quantum Information, Faculty of Physics and Astronomy, Adam Mickiewicz University, 61-614 Pozna\'n, Poland}

\author{Neill Lambert}
\affiliation{RIKEN Center for Quantum Computing, RIKEN, Wakoshi, Saitama 351-0198, Japan}

\author{Guang-Yin Chen}
\affiliation{Department of Physics, National Chung Hsing University, Taichung 402202, Taiwan}

\author{Franco Nori}
\email{fnori@riken.jp}
\affiliation{RIKEN Center for Quantum Computing, RIKEN, Wakoshi, Saitama 351-0198, Japan}
\affiliation{Physics Department, The University of Michigan, Ann Arbor, Michigan 48109-1040, USA.}

\author{Yueh-Nan Chen}
\email{yuehnan@mail.ncku.edu.tw}
\affiliation{Department of Physics, National Cheng Kung University, Tainan 701401, Taiwan}
\affiliation{Center for Quantum Frontiers of Research and Technology (QFort), Tainan 701401, Taiwan}
\affiliation{Physics Division, National Center for Theoretical Sciences, Taipei 106319, Taiwan}

\begin{abstract}
Quantum information scrambling (QIS) describes the rapid spread of initially localized information across an entire quantum many-body system through entanglement generation. Once scrambled, the original local information becomes encoded globally, inaccessible from any single subsystem. In this work, we introduce a circuit-based decoding protocol. By utilizing the concept of postselected closed timelike curves (PCTCs), we demonstrate how postselection allows us to interpret an ordinary quantum experiment as an example of a paradox-free trajectory, simulating a consistent time loop and reliable information recovery. Specifically, when conditioned on a final postselected outcome, this experiment can be interpreted as decoding the scrambled information even before the original information is generated. Furthermore, the success probability of the PCTC is governed by out-of-time-ordered correlations, which is a standard measure of QIS. We experimentally implement our protocol on cloud-based Quantinuum and IBM quantum processors. Our approach illuminates a unique quantum task under postselection: the causally consistent simulation of future-to-past scrambled information retrieval.
\end{abstract}

\maketitle

\section{Introduction}
Quantum information scrambling (QIS) arises from strong interactions in a many-body system~\cite{xu2024,swingle2018,iyoda2018}. Such dynamics rapidly correlates local information and generates many-body entanglement across all possible degrees of freedom within the global system~\cite{iyoda2018,ding2016,yungerhalpern2018,pappalardi2018,touil2020,lin2021,sharma2021,harris2022,zhu2022,lomonaco2025,Garcia2023,hosur2016,seshadri2018}. This delocalization process in many-body physics is typically characterized by the decay of the out-of-time-order correlator (OTOC)~\cite{xu2024,swingle2018,Fujii2025} and has sparked growing interest in various fields, such as quantum chaos~\cite{hosur2016,seshadri2018,dowling2023,juan2016,nahum2017,keyserlingk2018,Cotler2017,fan2017,Gu2017,khemani2018,zonnios2022,tezuka2023,bin2023,alonso2022}, black hole thermalization and information problems~\cite{hayden2007,sekino2008,Lashkari2013,Shenker2014,roberts2015,blake2016,Gao2017,Maldacena2017,chen2022b,liu2020,rinaldi2022,nation2012a}, quantum collision models~\cite{li2020,li2022,tian2024}, and quantum computing and error correction~\cite{shen2020,choi2020,li2025}.

Local information can be encoded into a global system by utilizing QIS. In a scrambled system, the information becomes dispersed and inaccessible via local measurements due to many-body entanglement. The robustness of recovering the original quantum information depends on the strength and structure of this entanglement~\cite{shen2020,yan2020}. Remarkably, it is possible to recover the information even when parts of the system are damaged or lost. For instance, if the scrambled state is damaged by a local measurement but this damage does not affect the useful information encoded in the entanglement correlations, one can still recover the original information~\cite{yan2020,choi2020}.

Another example is when a quantum state is thrown into a black hole, characterized by strong scrambling dynamics, and the information is believed to irretrievably disappear once it crosses the event horizon~\cite{hayden2007}. However, Ref.~\cite{hayden2007} proposed a thought experiment suggesting that it might be possible to probabilistically reconstruct the lost information from several emitted Hawking radiation photons from the black hole through a suitable decoding process. An efficient Yoshida-Kitaev probabilistic decoding protocol for this thought experiment was proposed~\cite{yoshida2017}, with some extended scenarios~\cite{yoshida2019,bao2021,schuster2022} experimentally demonstrated~\cite{landsman2019,blok2021,wang2022,shapoval2023}. Such a decoding protocol can also be considered postselected quantum teleportation. The preparation of Einstein-Podolsky-Rosen (EPR) pairs and the postselection of the measurement outcomes are necessary for the original state (information) to be successfully teleported (decoded) in their protocol.

While the Yoshida-Kitaev probabilistic decoding protocol has been widely adopted to characterize QIS, the physical intuition behind its essential ingredients, particularly the use of EPR pair preparation and postselection, has remained unclear. In this work, we clarify their role by drawing a parallel with the concept of postselected closed timelike curves (PCTCs)~\cite{Lloyd2011, Lloyd2011-2}.

A closed timelike curve (CTC) is a hypothetical concept in general relativity that describes a worldline in spacetime looping back on itself, allowing a particle to return to and interact with its own past~\cite{vanStockum1938,Godel1949}. This idea suggests the possibility of ``time travel'' and may raise some paradoxes~\cite{Morris1988}.
The PCTCs enable a specific formulation that combines CTCs with quantum mechanics based on quantum teleportation and postselection~\cite{Lloyd2011, Lloyd2011-2}, which can be simulated on quantum processors~\cite{buluta2009,georgescu2014,nation2012}. The postselection step introduces nonlinear effects that enforce the Novikov self-consistency principle~\cite{Novikov1990}, which dictates that only logically self-consistent events can occur. Paradoxical outcomes are assigned zero probability, and the success probability of time travel corresponds to the conditional probability of the remaining consistent events. Beyond foundational insights into quantum gravity~\cite{Lloyd2014,Xu2019,Kim2023}, the PCTC framework has found applications in quantum information processing~\cite{Oreshkov2015,Korotaev2015,Chiribella2022,Kiktenko2023} and quantum metrology~\cite{Nicole2023, Murch2024}.

In this work, we propose a PCTC-inspired protocol to transmit scrambled (encrypted) quantum information into the past. Importantly, our approach does not realize physical ``time travel''. It rather provides a controlled quantum simulation of the logical structure associated with PCTC models. We show that our protocol is operationally equivalent to the Yoshida-Kitaev probabilistic decoding protocol~\cite{yoshida2017,schuster2022}. Within our formulation, the roles of the EPR pair and postselection become physically transparent: They serve as the essential ingredients for constructing a paradox-free PCTC~\cite{Lloyd2011, Lloyd2011-2,Novikov1990}, thereby enabling consistent simulation
of successful future-to-past information recovery.

The protocol is quantified by the fidelity between the original prepared state and the decoded state. This decoding fidelity relies on the success probability of ``time travel'' through a PCTC, which in turn depends on the strength of QIS (characterized by the decay of the average value of the OTOC). Consequently, if the scrambling dynamics are sufficiently strong, the original information can be perfectly decoded (with unity fidelity). Finally, we demonstrate our decoding protocol using a four-qubit quantum circuit and then implement it on both Quantinuum~\cite{Quantinuum} and IBM quantum (IBMQ)~\cite{IBMQ,Qiskit} processors.

The rest of the paper is organized as follows. In Sec.~\ref{sec:PCTC-and-Protocol}, we review the concept of PCTCs and proceed to introduce our decoding protocol. In Sec.~\ref{sec:QIS}, we analyze the decay of the OTOC to characterize QIS and explore its connection to PCTCs within our decoding protocol. Section~\ref{sec:Experiment} presents the experimental demonstration of the protocol on the Quantinuum and IBMQ processors. Finally, we draw our conclusions in Sec.~\ref{sec:Conclusions}.

\section{Recovering scrambled information via postselected closed timelike curves}\label{sec:PCTC-and-Protocol}

\begin{figure}[!h]
    \centering
    \includegraphics[width=\linewidth]{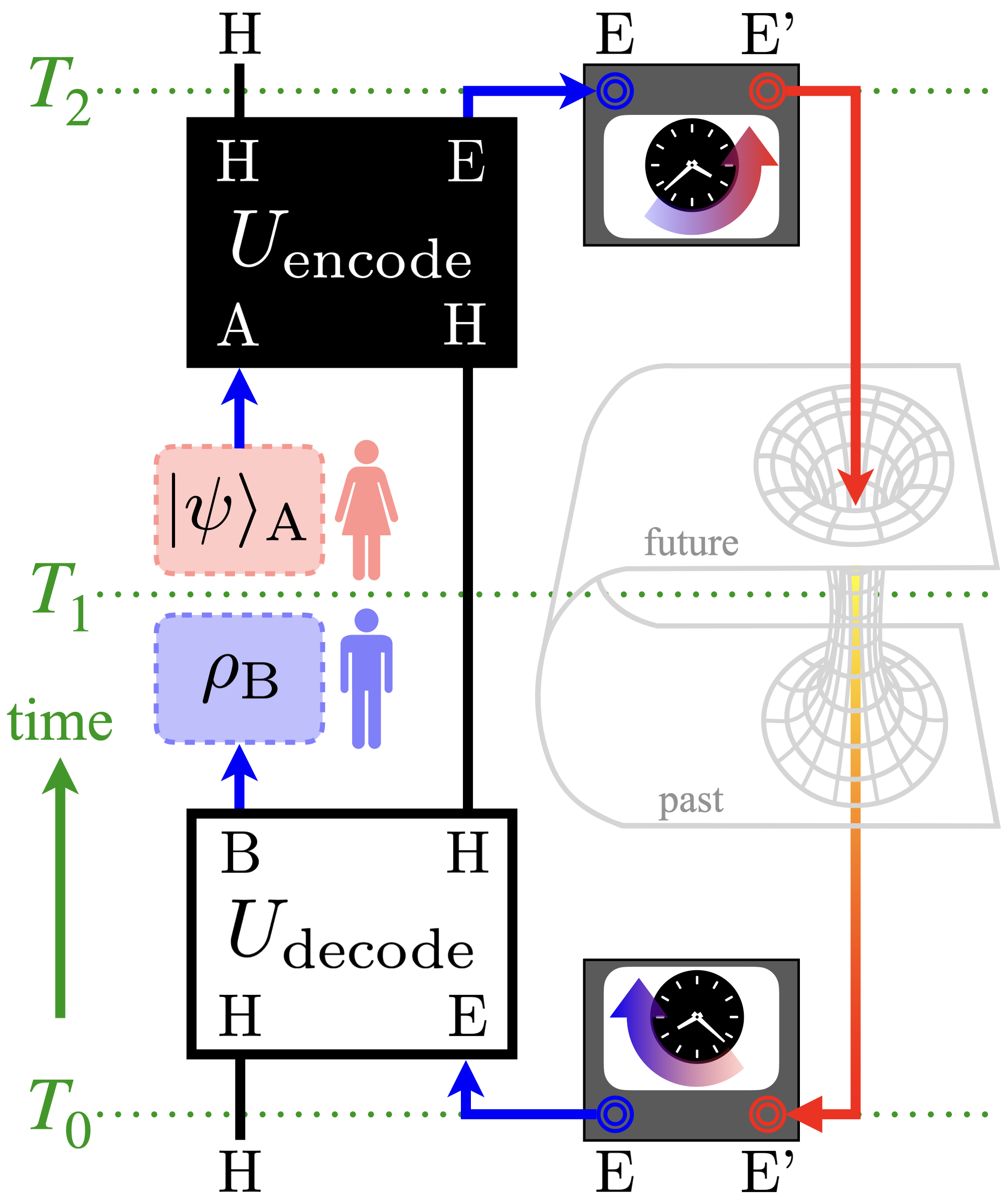}
    \caption{Schematic illustration of the decoding protocol. The black solid lines depict the timeline of the many-body system H going forward. The blue (red) arrows depict the timeline of a portion of the encoded information E (E') going forward (backward). Starting at time $T_1$, Alice (A) sends the information $\ket{\psi}$ into the scrambling unitary operator (scrambler) $U_\textrm{encode}=U_{\textrm{AH}\rightarrow\textrm{HE}}$, encoding the information in the joint system HE at $T_2$. By projecting the state of joint system EE' into an EPR state, the temporal direction for E is reversed, causing it to travel backward in time from $T_2$ to $T_0$. Because the state of system E and E' is initialized as an EPR state at $T_0$, the temporal direction for E' is reversed again. The decoding unitary operator $U_\textrm{decode}=U^\dagger_{\textrm{HE}\rightarrow\textrm{BH}}$ allows Bob (B) to decode Alice's information solely from the portion of the encoded information carried in E at $T_0$. Finally, Bob performs measurements before Alice prepares $\ket{\psi}$ at $T_1$, and the decoded state $\rho$ is conditioned on the postselected outcome obtained at $T_2$. The information is perfectly decoded if and only if the fidelity between $|\psi\rangle$ and $\rho$ is unity.}
    \label{fig:scheme}
\end{figure}

In this section, we first review the concept of PCTCs. Next, we introduce a protocol for recovering scrambled information from the future using PCTCs, as illustrated schematically in Fig.~\ref{fig:scheme}. Our protocol includes three main steps: (1) encoding quantum information using QIS, (2) simulating ``time travel'' via a PCTC to send part of the encoded information into the past, and (3) decoding the original information conditioned on the final postselected outcome from the PCTC.

\subsection{Postselected closed timelike curves}\label{sec:PCTC}
\begin{figure}[!h]
    \centering
    \includegraphics[width=0.92\linewidth]{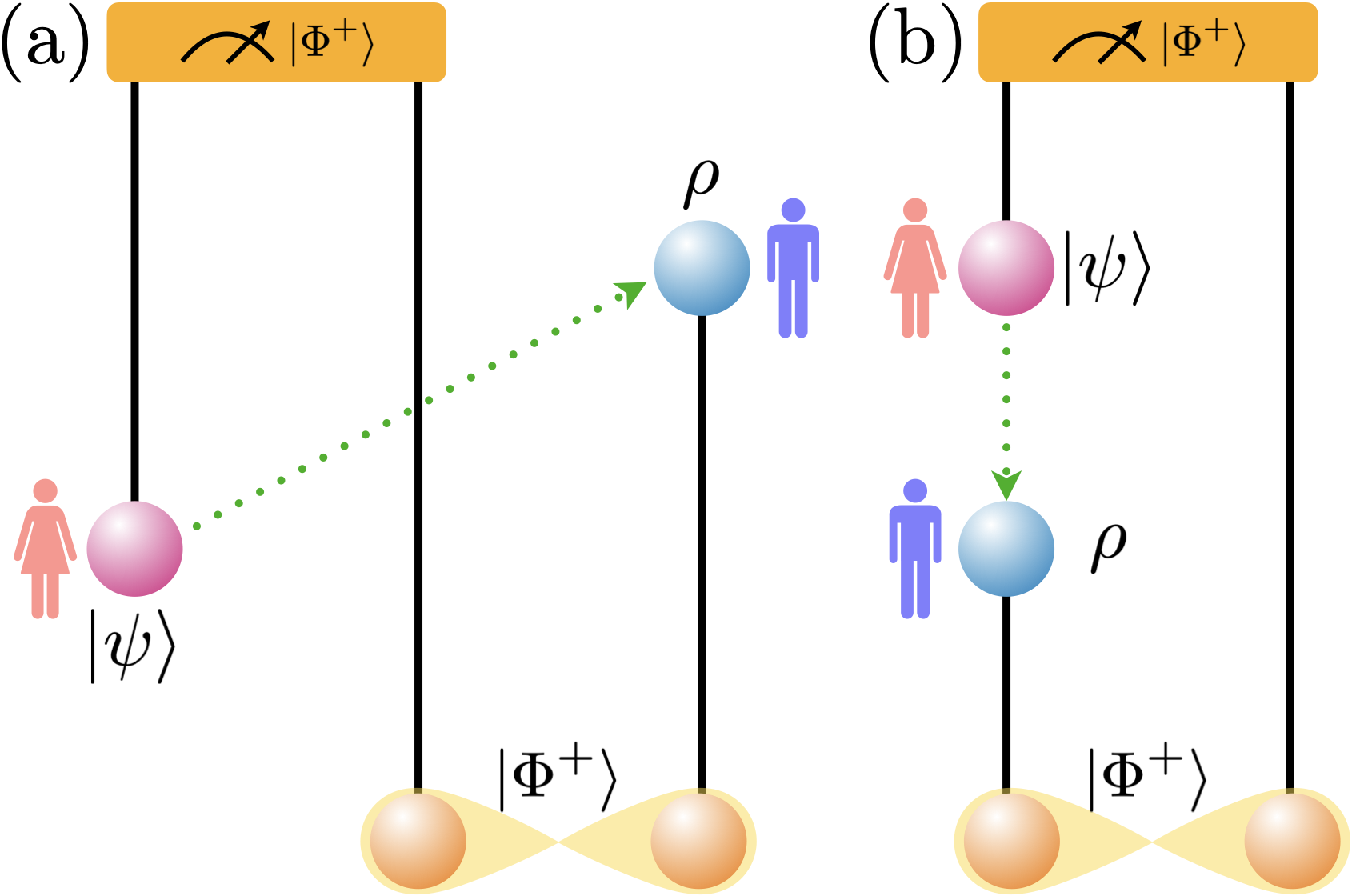}
    \caption{Comparison between postselected quantum teleportation and postselected closed timelike curves. (a) Postselected quantum teleportation. (b) Postselected closed timelike curves. Both scenarios rely on preparing an entangled pair $|\Phi^{+}\rangle$ and a joint measurement with postselection on the outcome corresponding to the state $|\Phi^{+}\rangle$. In the ideal case, the receiver obtains $\rho = |\psi\rangle\langle\psi|$, which is identical to the originally prepared state $|\psi\rangle$ in both scenarios.}
    \label{fig:PQT-and-PCTC}
\end{figure}

In this subsection, we recall the concept of PCTCs~\cite{Lloyd2011, Lloyd2011-2} by comparing it with postselected quantum teleportation. Figure~\ref{fig:PQT-and-PCTC}(a) illustrates the standard postselected quantum teleportation. Alice holds an unknown quantum state $|\psi\rangle$ and shares an EPR pair with Bob. After Alice performs a joint measurement and postselects the corresponding outcome, the state is teleported to Bob. The dotted green line indicates the instantaneous transfer of the quantum state under this successful postselection event. In this case, the teleportation succeeds without requiring classical communication because the postselection enforces the particular outcome.

Figure~\ref{fig:PQT-and-PCTC}(b) illustrates the same teleportation structure but interpreted as a PCTC~\cite{Lloyd2011, Lloyd2011-2}. Here, the postselection is depicted in a way that suggests a closed temporal loop. Under the PCTC interpretation, the postselection introduces nonlinear effects that enforce the Novikov self-consistency principle~\cite{Novikov1990}. The dotted green arrow represents the effective ``return'' of the system to an earlier time.

Thus, Fig.~\ref{fig:PQT-and-PCTC} highlights that the PCTC behavior emerges entirely from the same ingredients as in postselected quantum teleportation: an initial EPR pair and a final postselection on the same entangled state. When we select this measurement outcome, the effective evolution behaves as if part of the system had traveled backward in time.

\subsection{Decoding protocol}\label{sec:Protocol}
In this subsection, we introduce our decoding protocol, as shown in Fig.~\ref{fig:scheme}. We begin with Alice (A), who prepares an arbitrary pure quantum state $|\psi\rangle$ with Hilbert space dimension $d_{\textrm{A}}$ at time $T_1$. As time progresses from $T_1$ to $T_2$, system A interacts with a chronology-respecting many-body system H, and the global evolution of this interaction is described by an encoding unitary operator $U_\textrm{encode}=U_{\textrm{AH}\rightarrow\textrm{HE}}$. Here, $U$ can also be called a \textit{scrambler}, and the subscript labels the input subsystems (A and H) and output subsystems (H and E) of the operator. Note that the state of system H at $T_1$ will be discussed later. The quantum information $|\psi\rangle$ has become scrambled within the many-body entanglement of the joint system HE at $T_2$. Here, the output system E represents a part of the encoded information with its Hilbert space dimension also equal to $d_{\textrm{A}}$.

Next, the temporal direction of system E is reversed at time $T_2$, allowing it to travel back to $T_0$ (a time point before $T_1$) and then resume traveling forward in time. This is achieved by utilizing a PCTC, where we introduce a chronology-violating system E', depicted as the red curve in Fig.~\ref{fig:scheme}. If the initial state of E and E' at $T_0$ is a maximally entangled state, E' can be interpreted as a mirror of E that travels backward in time~\cite{Lloyd2011,Lloyd2011-2}. Furthermore, it is necessary to perform a projective (or selective) measurement on the joint system EE' at $T_2$, and the measurement outcome must remain consistent with itself (the same state) at $T_0$ to obey the Novikov self-consistency principle~\cite{Novikov1990}. Throughout this work, and without loss of generality, we consider the state of E and E' at $T_0$ to be prepared as an EPR pair~\cite{Neilsen2011}, namely,
\begin{equation}\label{eq:EPR}
|\textrm{EPR}\rangle_{\textrm{EE'}}=\frac{1}{\sqrt{d_{\textrm{A}}}}\sum_{i=0}^{d_{\textrm{A}}-1} |i\rangle_{\textrm{E}} \otimes |i\rangle_{\textrm{E'}}.
\end{equation}
Therefore, an EPR projective measurement must be performed at $T_2$, ensuring that the state of system E effectively travels backward to $T_0$. We denote the success probability of time travel through the PCTC (preparing and measuring $|\textrm{EPR}\rangle_{\textrm{EE'}}$ at $T_0$ and $T_2$, respectively) as $\mathcal{P}(\psi)$ for different state $|\psi\rangle$. 

At time $T_0$, a decoding operation $U_\textrm{decode}=U^\dagger_{\textrm{HE}\rightarrow\textrm{BH}}$ is applied to the joint system HE, and then Bob (B) subsequently receives a state $\rho$ (generally represented as a density matrix) before $T_1$. To demonstrate that the final decoding process requires only $U^\dagger_{\textrm{HE}\rightarrow\textrm{BH}}$ and the partially encoded information stored in E, we assume that the initial state of system H at $T_0$ is maximally mixed, which does not provide any additional information for the final decoding process. Finally, the fidelity between the states $|\psi\rangle$ and $\rho$ quantifies the entire decoding protocol, namely,
\begin{equation}\label{eq:F}
\mathcal{F}(\psi) = \langle \psi | \rho | \psi \rangle.
\end{equation}
The original information prepared by Alice is perfectly decoded if and only if the fidelity equals unity.

\begin{figure*}[ht]
    \centering
    \includegraphics[width=\linewidth]{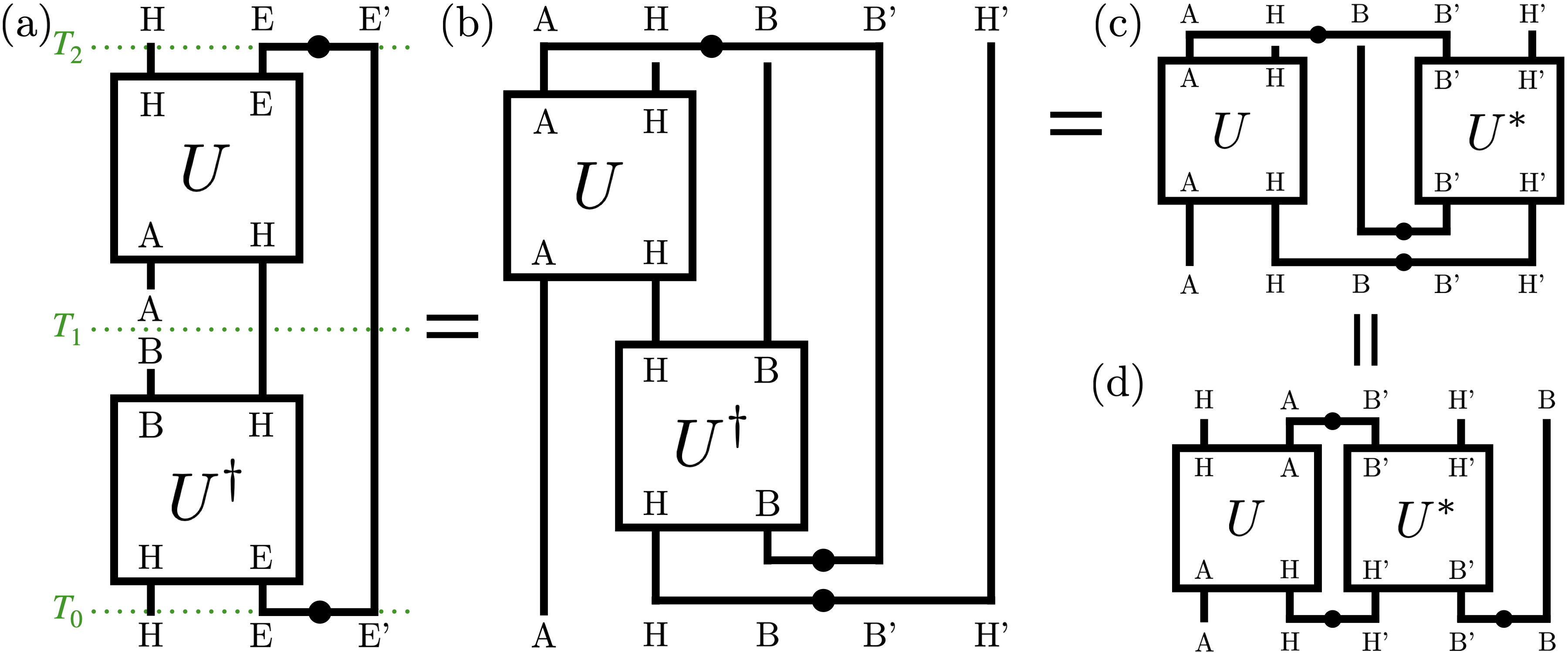}
    \caption{Proof of the equivalence of protocols using diagrammatic notation. (a) The diagram of our decoding protocol, where time progresses from $T_0$ to $T_2$ in the laboratory’s rest frame. Note that the black dot in the middle of the wire indicates a normalized EPR pair (scaled by $1/\sqrt{d}$). (b) We introduce the purification $|\textrm{EPR}\rangle_{\textrm{HH'}}$ for the initial state of the many-body system H, which is a maximally mixed state, with its reference system H'. For simplicity, we relabel the input system E (at time $T_0$), output system E (at time $T_2$), and system E' as B, A, and B', respectively. The equivalence between the diagrams in panels (b) and (c) follows from the properties of the EPR state, as described in Eq.~(\ref{seq:local_U_equality}). (d) Diagram obtained by rearranging the input and output Hilbert space order from the diagram in panel (c).}
    \label{fig:diagrammatic-proof}
\end{figure*}

Here, we prove that our decoding protocol is equivalent to the Yoshida-Kitaev probabilistic decoding protocol~\cite{yoshida2017}. The proof under the diagrammatic notation is presented in Fig.~\ref{fig:diagrammatic-proof}, and a similar discussion for another protocol can also be found in Ref.~\cite{schuster2022}.

In our protocol, the initial state of the many-body system H at $T_0$ is a maximally mixed state. However, the initial state in the Yoshida-Kitaev probabilistic decoding protocol is derived from an entangled EPR pair between system H and an auxiliary system H', i.e., $|\textrm{EPR}\rangle_{\textrm{HH'}}$. Incorporating this constraint into our protocol, the final unnormalized state $|f(\psi)\rangle$, corresponding to the input quantum information $|\psi\rangle$ from Alice, can be expressed as
\begin{align}\label{seq:final_ket_ours}
|f&(\psi)\rangle_{\textrm{BHH'}}\notag\\
&={}_{\textrm{AB'}}{\langle\textrm{EPR}}|U_{\textrm{AH}}U^\dagger_{\textrm{BH}} |\psi\rangle_\textrm{A}\otimes|\textrm{EPR}\rangle_\textrm{BB'}\otimes|\textrm{EPR}\rangle_\textrm{HH'},
\end{align}
as shown in Fig.~\ref{fig:diagrammatic-proof}(b). Here, without loss of generality, we relabel the input system E (at time $T_0$), output system E (at time $T_2$), and the chronology-violating system E' as B, A, and B', respectively. This relabeling is justified because, from the perspective of PCTCs, all these systems represent the same system that carries the information at different time points or travels in different directions in time. Additionally, for simplicity, we rearrange both the input and output Hilbert spaces into the same order as AHBB'H' in the diagrammatic notation, as shown in Fig.~\ref{fig:diagrammatic-proof}(b). Note that although the diagrammatic notation of $|\textrm{EPR}\rangle_\textrm{HH'}$ in Fig.~\ref{fig:diagrammatic-proof}(b) appears as a single wire, it actually represents multiple EPR pairs shared between the corresponding subsystems of H and H'.

A key element of this proof is the following identity, which holds when an arbitrary local unitary $U$ is applied to an EPR pair, namely,
\begin{equation}\label{seq:local_U_equality}
U_\textrm{X}\otimes\openone_\textrm{X'}|\textrm{EPR}\rangle_\textrm{XX'} = \openone_\textrm{X}\otimes U_\textrm{X'}^\textrm{T}|\textrm{EPR}\rangle_\textrm{XX'},
\end{equation}
where the superscript T represents the transpose of an operator and $\openone$ is the identity operator. Thus, using the equality given in Eq.~(\ref{seq:local_U_equality}), we can replace the decoding unitary operation $U^\dagger_{\textrm{BH}} \otimes \openone_{\textrm{B'H'}}$ with $\openone_{\textrm{BH}} \otimes U^*_{\textrm{B'H'}}$ in Eq.~(\ref{seq:final_ket_ours}), namely,
\begin{align}\label{seq:final_ket_Y_K}
|f&(\psi)\rangle_{\textrm{BHH'}}\notag\\
&={}_{\textrm{AB'}}{\langle\textrm{EPR}|}U_{\textrm{AH}}U^*_{\textrm{B'H'}} |\psi\rangle_\textrm{A}\otimes|\textrm{EPR}\rangle_\textrm{BB'}\otimes|\textrm{EPR}\rangle_\textrm{HH'},
\end{align}
for which the corresponding diagrammatic notation is given in Fig.~\ref{fig:diagrammatic-proof}(c). After rearranging the input Hilbert space order to AHH'B'B and the output Hilbert space order to HAB'H'B in the diagrammatic notation [as shown in Fig.~\ref{fig:diagrammatic-proof}(d)], the unnormalized state $|f(\psi)\rangle$ is identical to the one presented in the Yoshida-Kitaev probabilistic decoding protocol~\cite{yoshida2017}. Consequently, the normalized decoded state is given by
\begin{equation}\label{seq:final_rho}
\rho_{\textrm{B}} = \frac{\textrm{Tr}_{\textrm{HH'}}\left[|f(\psi)\rangle\langle f(\psi)|_{\textrm{BHH'}}\right]}{\mathcal{P}(\psi)},
\end{equation}
where the success probability $\mathcal{P}(\psi)$ of time travel is expressed as
\begin{equation}\label{seq:P}
\mathcal{P}(\psi)= \langle f(\psi) | f(\psi) \rangle.
\end{equation}

We emphasize that, in our protocol, the input state to the encoding scrambler $U$ for system H must be the output from the decoding operation $U^\dagger$. This indicates that the decoding operation must occur before the encoding scrambler, unlike in the Yoshida-Kitaev probabilistic decoding protocol. Furthermore, their protocol requires generating multiple EPR pairs depending on the size of both systems H and E, whereas our protocol uses fewer entanglement resources. For example, when considering an $n$-qubit scrambler and performing quantum state tomography on system B over $S$ shots, the Yoshida-Kitaev probabilistic decoding protocol requires a total of $(n-1)S$ physical EPR pairs. Because the success probability $\mathcal{P}(\psi)$ is identical in both protocols, they require the same sampling overhead ($S$) to achieve an equivalent number of successful postselection events. However, our protocol requires only a single physical EPR pair between systems E and E' per shot ($S$ pairs in total). The maximally mixed state of the remaining $(n-1)$ qubits in system H is simply achieved by preparing each qubit in either the $|0\rangle$ or $|1\rangle$ state with equal classical probability across the $S$ shots. Thus, for any $n > 2$, our protocol strictly reduces the total experimental entanglement cost by a factor of $(n-1)$.

Furthermore, the product of the success probability $\mathcal{P}(\psi)$ and the fidelity $\mathcal{F}(\psi)$ is lower bounded by $d_{\textrm{A}}^{-2}$~\cite{yoshida2017}, namely,
\begin{equation}\label{eq:PF}
\mathcal{P}(\psi)\mathcal{F}(\psi) \geq \frac{1}{d_{\textrm{A}}^2}~~~\forall~~\psi
\end{equation}
(see Appendix~\ref{sec:PF_lower_bound} for the derivation). This indicates that the original information is accurately decoded (with $\mathcal{F} \approx 1$) when the success probability is sufficiently low (with $\mathcal{P} \approx d_{\textrm{A}}^{-2}$) in the ideal case.

\section{Quantum information scrambling and out-of-time-order correlators}\label{sec:QIS}

Quantum information scrambling is typically characterized by the decay of the OTOC~\cite{xu2024,swingle2018,Fujii2025}, defined as
\begin{equation}\label{eq:OTOC}
\mathcal{O}(W,V)\equiv \langle W_\textrm{E}^\dagger(t) V_\textrm{A}^\dagger W_\textrm{E}(t)V_\textrm{A}\rangle,
\end{equation}
where $V_\textrm{A}$ and $W_\textrm{E}$ are initially commuting unitary and Hermitian operators acting on separate systems A and E, respectively. In our decoding protocol, any pure state $|\psi\rangle_\textrm{A}$ prepared by Alice can be written as $|\psi\rangle_\textrm{A} = V_\textrm{A} |\phi_0\rangle_\textrm{A}$, where $|\phi_0\rangle_\textrm{A}$ is a fixed reference state. Thus, $V_\textrm{A}$ represents a generic state-preparation operation performed solely on system A before the scrambling dynamics is applied. The operator $W_\textrm{E}$ can be considered as a local perturbation applied to E at time $T_2$, and $W_\textrm{E}(t)=U_{\textrm{decode}}W_\textrm{E}U_{\textrm{encode}}=U^\dagger W_\textrm{E} U$ is the time-evolved version of $W_\textrm{E}$ in the Heisenberg picture according to the scrambling unitary operator $U$. From the perspective of PCTCs, one can also interpret $W_{\textrm{E}}(t)$ as follows. After applying $U_{\textrm{encode}}$, a local perturbation $W_{\textrm{E}}$ is introduced on system E at $T_2$. The system E is then effectively propagated backward in time through the PCTC and decoded by $U_{\textrm{decode}}$. Thus, the action of $W_{\textrm{E}}(t)$ incorporates both the forward scrambling dynamics and the nontrivial temporal structure introduced by the PCTC, offering a physical intuition for how a local disturbance at $T_2$ influences the reconstructed information $\rho_\textrm{B}$ at an earlier time before $T_1$.

The quantity $\mathcal{O}(W,V)$ can be understood as the overlap between two processes: (1) $W_\textrm{E}(t)\, V_\textrm{A}|\phi_0\rangle_\textrm{A}$, in which $V_\textrm{A}$ is applied first and then the time-evolved perturbation $W_\textrm{E}(t)$ acts, and (2) $V_\textrm{A}W_\textrm{E}(t)|\phi_0\rangle_\textrm{A}$, in which $W_\textrm{E}(t)$ acts first followed by $V_\textrm{A}$. When $t=0$, the value of the OTOC $\mathcal{O}(W, V) = 1~\forall~W~\textrm{and}~V$ because $W(0)$ commutes with $V$. As scrambling progresses, $W(t)$ ceases to commute with $V$ as it becomes increasingly nonlocal, thereby leading to the rapid decay of the OTOC. Such a decay directly measures the delocalization of information initially localized in system A at $T_1$ across the joint system H and E at $T_2$.

The average value of the OTOC in Eq.~(\ref{eq:OTOC}) can be directly calculated by the success probability $\mathcal{P(\psi)}$ of ``time travel'' in our decoding protocol~\cite{yoshida2017}:
\begin{equation}\label{eq:avg-OTOC}
    \mathcal{O}_\avg = \iint dWdV~\mathcal{O}(W, V)=\int d\psi \mathcal{P}(\psi)
\end{equation}
(see Appendix~\ref{sec:QIS_PCTC_proof} for the derivation). Here, the double integral is defined with respect to the normalized Haar measure on the unitary group. This means that $dW$ and $dV$ denote sampling $W_\textrm{E}$ and $V_\textrm{A}$ uniformly from all unitary operators, in the sense that the distribution is unchanged under left or right multiplication by any unitary. Similarly, $d\psi$ is the normalized Haar measure over pure states $|\psi\rangle$ in our protocol. Equation~(\ref{eq:avg-OTOC}) shows that the average success probability of ``time travel'' through the PCTC decreases as the average value of the OTOC decays.

Furthermore, one can observe that the average value of the OTOC decays to $d_{\textrm{A}}^{-2}$ for the aforementioned perfect decoding case [$\mathcal{F}(\psi) = 1$ and $\mathcal{P}(\psi) = d_{\textrm{A}}^{-2}~\forall~\psi$]. This indicates that the unitary evolution $U$ must generate sufficiently strong scrambling dynamics (which makes $\mathcal{O}_\textrm{avg}$ decay to $d_{\textrm{A}}^{-2}$) so that the information $|\psi\rangle$ can be fully encoded through the many-body entanglement in the joint system HE at $T_2$, and then perfectly decoded.

\begin{figure}[tb]
    \centering
    \includegraphics[width=\linewidth]{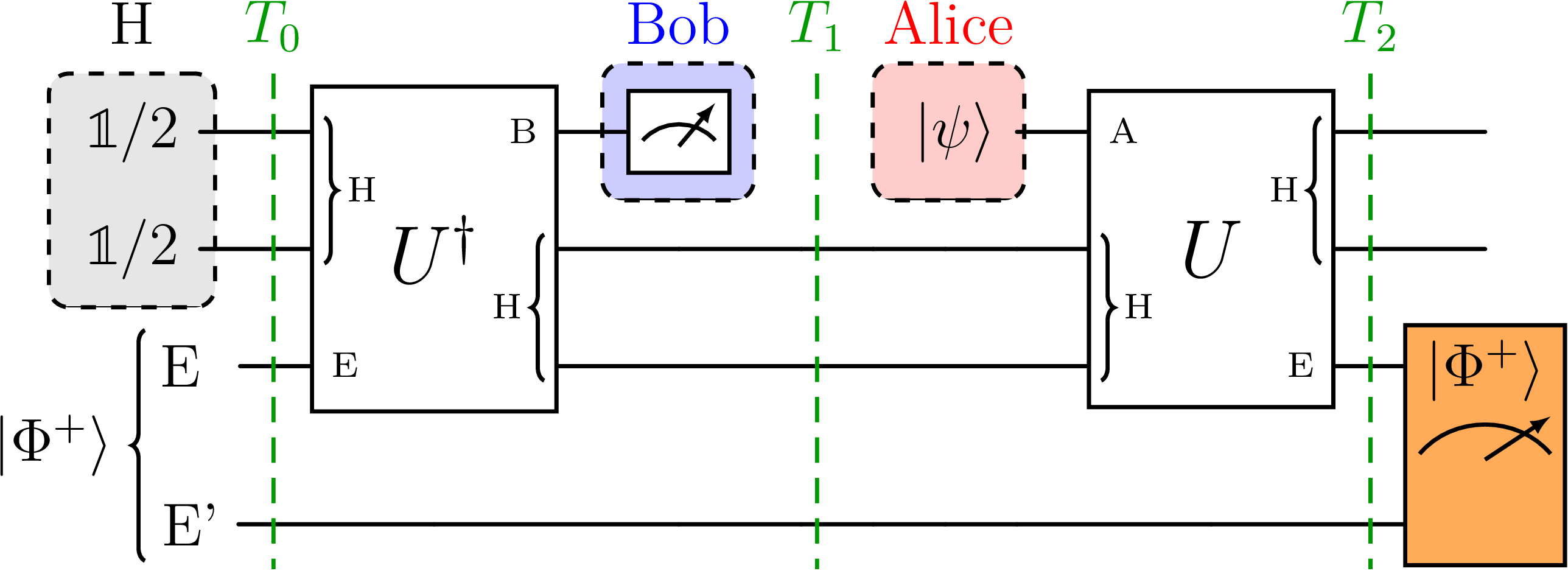}
    \caption{Experimental demonstration of the decoding protocol using a four-qubit quantum circuit. Time progresses from $T_0$ to $T_2$ in the laboratory's rest frame. Initially, the system H is prepared as a two-qubit maximally mixed state, while the systems E and E' are initialized as $|\Phi^+\rangle=(|00\rangle+|11\rangle)/\sqrt{2}$, which is one of the Bell states. The three-qubit decoding operation $U^\dagger$ is applied to decode the information stored solely in system E, and Bob performs quantum state tomography on qubit B before $T_1$. After $T_1$, Alice prepares the information $|\psi\rangle$ and encodes it using the scrambler $U$. To complete the decoding protocol, a Bell state measurement is performed after $T_2$, and Bob needs to postselect his results corresponding to the state $|\Phi^+\rangle$.}
    \label{fig:circuit}
\end{figure}

\section{Experimental demonstration}\label{sec:Experiment}
As we mentioned above, although the existence of CTCs is still vague, one can probabilistically simulate it based on postselected quantum teleportation. Here, we demonstrate our decoding protocol using a four-qubit quantum circuit, shown in Fig.~\ref{fig:circuit}. The circuit is implemented using Quantinuum~\cite{Quantinuum} and IBMQ~\cite{IBMQ,Qiskit} processors.

In the following, we explain the circuit in detail in the laboratory's rest frame (from $T_0$ to $T_2$). At time $T_0$, system H is initialized as a two-qubit maximally mixed state [$(\openone/2)\otimes (\openone/2)$], a uniform mixture of states in the computational basis. Thus, we uniformly prepare each of the two qubits in the state $|0\rangle$ or $|1\rangle$. Systems E and E' are initialized as a two-qubit EPR pair, as shown in Eq.~(\ref{eq:EPR}) with $d_{\textrm{A}}=2$:
\begin{equation}\label{eq:Bell-state}
|\Phi^+\rangle_{\textrm{EE'}}=\frac{1}{\sqrt{2}}\left(|0\rangle_\textrm{E}\otimes|0\rangle_\textrm{E'}+|1\rangle_\textrm{E}\otimes|1\rangle_\textrm{E'}\right),
\end{equation}
which is a Bell state~\cite{Neilsen2011}.

After applying the decoding operation $U^\dagger$ to the joint system HE, we perform quantum state tomography (i.e., single-qubit Pauli measurements~\cite{Neilsen2011}) on qubit B before $T_1$. To ensure the information is from the future, we reset qubit B and then prepare a pure quantum state $|\psi\rangle$ on the same qubit (which we rename qubit A after $T_1$). We emphasize that the state $|\psi\rangle$ is prepared freely, with no knowledge of the previous measurement results. We then apply the encoding scrambler $U$ to encode the information within the joint system HE at $T_2$. To complete the decoding protocol, a Bell state measurement is performed on the joint system EE' after $T_2$, and we postselect the outcome corresponding to $|\Phi^+\rangle_{\textrm{EE'}}$ as in Eq.~(\ref{eq:Bell-state}). Finally, we reconstruct the density matrix $\rho$ from the postselected measurement results and calculate the fidelity $\mathcal{F}(\psi)$, given in Eq.~(\ref{eq:F}), between the states $\rho$ and $|\psi\rangle$, confirming that the encoded quantum information has been successfully decoded.

To characterize the nature of the scrambling dynamics, we repeat this protocol by encoding the six different initial states and two different scramblers: $|\psi\rangle\in\{|x_-\rangle,|x_+\rangle,|y_-\rangle,|y_+\rangle,|z_-\rangle,|z_+\rangle\}$, where $
|x_\pm\rangle\equiv\frac{1}{\sqrt{2}}(|0\rangle\pm|1\rangle)$, $|y_\pm\rangle\equiv\frac{1}{\sqrt{2}}(|0\rangle\pm i|1\rangle)$, $|z_-\rangle\equiv|1\rangle$, and $|z_+\rangle\equiv|0\rangle$. For the scrambling unitary operator $U$, we consider the three-qubit Clifford scramblers proposed in Ref.~\cite{yoshida2019}. The first Clifford scrambler, denoted as $U_\textrm{q}$, is capable of fully delocalizing arbitrary quantum information. The explicit matrix form of this unitary operator is given by
\begin{equation}\label{eq:U_q}
U_\textrm{q} = \frac{1}{2\sqrt{2}}\begin{pmatrix}
    1 & 1 & 1 & -1 & 1 & -1 & -1 & -1\\
    1 & -1 & 1 & 1 & 1 & 1 & -1 & 1\\
    1 & 1 & -1 & 1 & 1 & -1 & 1 & 1\\
    -1 & 1 & 1 & 1 & -1 & -1 & -1 & 1\\
    1 & 1 & 1 & -1 & -1 & 1 & 1 & 1\\
    -1 & 1 & -1 & -1 & 1 & 1 & -1 & 1\\
    -1 & -1 & 1 & -1 & 1 & -1 & 1 & 1\\
    -1 & 1 & 1 & 1 & 1 & 1 & 1 & -1
\end{pmatrix}.
\end{equation}
The success probability for an arbitrary state $|\psi\rangle$ under the scrambler $U_\textrm{q}$ is $\mathcal{P}(\psi)=0.25$. According to Eq.~(\ref{eq:PF}), this leads to an estimated fidelity of unity for any state $|\psi\rangle$, indicating perfect decoding. 

To show the strength of $U_\textrm{q}$, we compare it with another Clifford scrambler, denoted as $U_\textrm{c}$, which only scrambles classical information:
\begin{equation}\label{eq:U_c}
U_\textrm{c} = \textrm{diag}(1, 1, 1, -1, 1, -1, -1, -1).
\end{equation}
In this case, the average success probability of the PCTC $\mathcal{P}_\avg=0.5$, resulting in a lower bound on the fidelity of $0.5$. Only states $|\psi\rangle$ prepared in $|0\rangle$ or $|1\rangle$ can be perfectly decoded, while the fidelity of all other states is $0.5$. The circuit decompositions for both Clifford scramblers ($U_\textrm{q}$ and $U_\textrm{c}$) are given in Appendix~\ref{sec:Circuit_decomposition}.

We performed the experiments using the H1-1 processor from
Quantinuum~\cite{Quantinuum} and the ibm\_torino processor from IBMQ~\cite{IBMQ,Qiskit}. The topology and calibration data are also presented in Appendix~\ref{sec:Circuit_decomposition}. The experimental results, as shown in Table~\ref{tab:exp-result}, are obtained through $4000$ and $40000$ shots for each measurement procedure in the state tomography using the H1-1 and ibm\_torino, respectively. For both processors, when $U_\textrm{q}$ is employed as the scrambler in the protocol, we observe that the success probability of ``time travel'' $\mathcal{P}(\psi)$ approximately approaches $d_{\textrm{A}}^{-2}=0.25$ for all initial states $|\psi\rangle$, which also means that $\mathcal{O}_{\textrm{avg}}\approx0.25$ due to Eq.~(\ref{eq:avg-OTOC}). Due to the strong QIS $U_\textrm{q}$, the decoding protocol succeeds with an average fidelity of $0.985$ ($0.8453$) using H1-1 (ibm\_torino). Additionally, it can be observed that the average fidelity achieved on the H1-1 processor is higher than that on ibm\_torino. This is attributed to the H1-1 processor's longer qubit relaxation and decoherence times, as well as lower readout and gate error rates (see Appendix~\ref{sec:Circuit_decomposition} for the calibration data of both processors).

In contrast, applying the same protocol with the scrambler $U_\textrm{c}$ using ibm\_torino results in a low decoding fidelity. Although the fidelity for states $|z_\pm\rangle$ remains relatively high (around $0.91$), the fidelity drops significantly to around $0.5$ for other states ($|x_\pm\rangle$ and $|y_\pm\rangle$). This indicates that $U_\textrm{c}$ only has the capability to scramble classical information ($|z_\pm\rangle$) instead of general quantum information. The large value of $\mathcal{O}_\textrm{avg}=\mathcal{P}_\textrm{avg}=0.4433$ further supports this conclusion.

\begin{table}[tb]
    \centering
    \caption{Experimental results of the decoding fidelity ($\mathcal{F}$) and the success probability of ``time travel'' ($\mathcal{P}$) for different initial states $|\psi\rangle$ under Clifford scramblers ($U_\textrm{q}$ and $U_\textrm{c}$). As shown in Eq.~(\ref{eq:avg-OTOC}), the average value of the out-of-time-order correlator ($\mathcal{O}_{\textrm{avg}}$) equals the average value of $\mathcal{P}(\psi)$. Note that the results obtained from IBMQ can be substantially improved (reaching fidelities comparable to those reported by Quantinuum) through the use of advanced error mitigation techniques developed by Algorithmiq~\cite{algorithmiq-private}.}\label{tab:exp-result}
    \begin{tabular}{cccccccccc}
    \hline
    \hline
    \multirow{2}{*}{$|\psi\rangle$} && \multicolumn{2}{c}{$U_\textrm{q}$ (Quantinuum)} && \multicolumn{2}{c}{$U_\textrm{q}$ (IBMQ)} && \multicolumn{2}{c}{$U_\textrm{c}$ (IBMQ)}\\
    \cline{3-4}\cline{6-7}\cline{9-10}
    && $\mathcal{F(\psi)}$ & $\mathcal{P(\psi)}$ && $\mathcal{F(\psi)}$ ~&~ $\mathcal{P(\psi)}$ && $\mathcal{F(\psi)}$ ~&~ $\mathcal{P(\psi)}$\\
    \hline
    $|x_-\rangle$ && 0.976 & 0.258 && 0.8320 & 0.2519 && 0.5021 & 0.4430\\
    $|x_+\rangle$ && 0.986 & 0.249 && 0.8219 & 0.2619 && 0.5014 & 0.4440\\
    $|y_-\rangle$ && 0.990 & 0.243 && 0.8681 & 0.2592 && 0.4949 & 0.4454\\
    $|y_+\rangle$ && 0.988 & 0.256 && 0.8506 & 0.2578 && 0.5082 & 0.4406\\
    $|z_-\rangle$ && 0.983 & 0.253 && 0.8479 & 0.2566 && 0.9092 & 0.4405\\
    $|z_+\rangle$ && 0.987 & 0.252 && 0.8512 & 0.2564 && 0.9130 & 0.4461\\
    \hline
    Average && 0.985 & 0.252 && 0.8453 & 0.2573 && 0.6381 & 0.4433\\
    \hline
    \hline
    \end{tabular}
\end{table}

\section{Conclusions}\label{sec:Conclusions}
We have proposed a protocol and experimentally demonstrated it on both Quantinuum and IBM quantum processors, leveraging QIS and PCTCs to decode encrypted quantum information from the future into the past. Although QIS delocalizes the original quantum information through many-body entanglement, we can still retrieve it by sending part of the scrambled information into the past. Here, the ``time travel'' is simulated through PCTCs, and its average success probability is directly related to the average value of OTOC in QIS. Notably, we show that the choice of the postselected outcome in our protocol is governed by the Novikov self-consistency principle: The outcome must be consistent with the state in which it was prepared in the past. Furthermore, we observe that the perfect decoding fidelity of our protocol requires strong QIS, which corresponds to a sufficiently low average value of the OTOC.

To substantiate our theoretical framework, we conducted a proof-of-principle experiment, highlighting the essential roles that QIS and PCTCs play in the success of decoding an encrypted message in our protocol. Importantly, our protocol preserves the causality of information~\cite{Pawowski2009}. Sending a secret to the past via PCTCs is like ouroboros, a serpent devouring its own tail. Any attempt to alter the past is a futile endeavor because whatever happened, happened.

Our work paves the way for several intriguing future research directions. Our protocol allows the estimation of the average value of the OTOC for arbitrary QIS using only one additional entangled EPR pair. While our demonstration utilizes a three-qubit scrambler, exploring more complex scrambling dynamics could enhance the understanding of the relationship between the decoding fidelity and different types of QIS, potentially optimizing the protocol further. Furthermore, our protocol establishes a connection between QIS and PCTCs, exploring insights related to black holes, wormholes, and ``time travel.''

\section*{Acknowledgments}
The authors acknowledge fruitful discussions with Chia-Yi Ju, Gelo Noel Tabia, and members of the Algorithmiq team: Keijo Korhonen, Matteo Rossi, Sergei Filippov, and Sabrina Maniscalco. The authors also acknowledge the Cloud Computing Center for Quantum Science \& Technology at NCKU (NSTC Grant No. 114-2119-M-006-003), NTU-IBM Q Hub, and IBM Quantum for providing them a platform to implement the experiments.
Y.-T.H. acknowledges the support of the National Science and Technology Council, Taiwan (NSTC Grant No. 113-2917-I-006-024).
A.M. was supported by the Polish National Science Centre (NCN) under the Maestro Grant No. DEC-2019/34/A/ST2/00081.
N.L. was supported by MEXT KAKENHI (Grant No. JP24H00816 and Grant No. JP24H00820).
G.-Y.C. acknowledges the support of the National Science and Technology Council, Taiwan (NSTC Grant No. 113-2112-M-005-008).
F.N. is supported in part by the Japan Science and Technology Agency (JST) [via the CREST Quantum Frontiers program Grant No. JPMJCR24I2, the Quantum Leap Flagship Program (Q-LEAP), the Moonshot R\&D Grant Number JPMJMS256E, and the ASPIRE program (Grant Number JPMJAP2513)], and the Office of Naval Research (ONR) Global (via Grant No. N62909-23-1-2074).
Y.-N.C. acknowledges the support of the National Center for Theoretical Sciences and the National Science and Technology Council, Taiwan (NSTC Grant No. 114-2112-M-006-015-MY3).

\appendix

\section{A lower bound for the product of the PCTC success probability and decoding fidelity}\label{sec:PF_lower_bound}

We demonstrate that the product of the PCTC success probability [$\mathcal{P}$ in Eq.~(\ref{seq:P})] and the decoding fidelity [$\mathcal{F}$ in Eq.~(\ref{eq:F})] in our decoding protocol is lower bounded by $d_{\textrm{A}}^{-2}$ for an arbitrary encoded state $|\psi\rangle$. Here, $d_{\textrm{A}}$ denotes the Hilbert space dimension of the original input quantum information $|\psi\rangle$ from Alice. Note that a similar proof can also be found in Ref.~\cite{yoshida2017}. We begin by substituting Eq.~(\ref{seq:final_rho}) into Eq.~(\ref{eq:F}) and multiplying by $\mathcal{P}(\psi)$, namely,
\begin{align}\label{seq:PF_1}
\mathcal{P}&(\psi)\mathcal{F}(\psi)\notag\\
&={}_{\textrm{B}}{\langle\psi|}\Big(\textrm{Tr}_{\textrm{HH'}}\left[|f(\psi)\rangle\langle f(\psi)|_{\textrm{BHH'}}\right]\Big)|\psi\rangle_{\textrm{B}}\notag\\
&\geq {}_{\textrm{B}}{\langle\psi|}\Big(\textrm{Tr}_{\textrm{HH'}}\left[|\textrm{EPR}\rangle\langle\textrm{EPR}|_{\textrm{HH'}}\times|f(\psi)\rangle\langle f(\psi)|_{\textrm{BHH'}}\right]\Big)|\psi\rangle_{\textrm{B}}\notag\\
&={}_{\textrm{B}}{\langle\psi|}\otimes{}_{\textrm{HH'}}{\langle\textrm{EPR}|} \Big(|f(\psi)\rangle\langle f(\psi)|_{\textrm{BHH'}}\Big) |\psi\rangle_{\textrm{B}} \otimes |\textrm{EPR}\rangle_{\textrm{HH'}}\notag\\
&=\Big|{}_{{\textrm{B}}}{\langle\psi|}\otimes{}_{\textrm{HH'}}{\langle\textrm{EPR}|}\times|f(\psi)\rangle_{\textrm{BHH'}}\Big|^2.
\end{align}
The above inequality holds because we also perform an EPR projective measurement on the joint system HH' before tracing them out. Next, by substituting Eq.~(\ref{seq:final_ket_Y_K}) into Eq.~(\ref{seq:PF_1}), one obtains
\begin{widetext}
\begin{equation}\label{seq:PF_2}
\mathcal{P}(\psi)\mathcal{F}(\psi)\geq\Big|{}_{\textrm{B}}{\langle\psi|}\otimes{}_{\textrm{AB'}}{\langle\textrm{EPR}|}\otimes{}_{\textrm{HH'}}{\langle\textrm{EPR}|}U_{\textrm{AH}}U^*_{\textrm{B'H'}} |\psi\rangle_\textrm{A}\otimes|\textrm{EPR}\rangle_\textrm{BB'}\otimes|\textrm{EPR}\rangle_\textrm{HH'}\Big|^2.
\end{equation}
\end{widetext}
Note that $U_{\textrm{AH}}$ and $U^*_{\textrm{B'H'}}$ commute, and we can apply the equality given in Eq.~(\ref{seq:local_U_equality}) once more to replace $U^*_{\textrm{B'H'}}$ with $U^\dagger_{\textrm{AH}}$ since
\begin{align}
{}_{\textrm{AB'}}{\langle\textrm{EPR}|}&\otimes{}_{\textrm{HH'}}{\langle\textrm{EPR}|}\openone_{\textrm{AH}}\otimes U^*_{\textrm{B'H'}} \notag\\
&= {}_{\textrm{AB'}}{\langle\textrm{EPR}|}\otimes{}_{\textrm{HH'}}{\langle\textrm{EPR}|}U^\dagger_{\textrm{AH}}\otimes \openone_{\textrm{B'H'}}.
\end{align}
Thus, with $U^\dagger_{\textrm{AH}}U_{\textrm{AH}} = \openone_{\textrm{AH}}$, Eq.~(\ref{seq:PF_2}) results in
\begin{align}\label{seq:PF_3}
\mathcal{P}(\psi)\mathcal{F}(\psi)&\geq\Big|{}_{{\textrm{B}}}{\langle\psi|}\otimes{}_{\textrm{AB'}}{\langle\textrm{EPR}|}\times|\psi\rangle_\textrm{A}\otimes|\textrm{EPR}\rangle_\textrm{BB'}\Big|^2\notag\\
&=\Big|\frac{1}{\sqrt{d_{\textrm{A}}}}\frac{1}{\sqrt{d_{\textrm{A}}}}\sum_{i=0}^{d_{\textrm{A}}-1}\sum_{j=0}^{d_{\textrm{A}}-1}\langle i|\psi \rangle_{\textrm{A}} \times \langle \psi|j \rangle_{\textrm{B}} \times \langle i|j \rangle_{\textrm{B'}}\Big|^2\notag\\
&=\Big|\frac{1}{d_{\textrm{A}}}\sum_{i=0}^{d_{\textrm{A}}-1}\langle i|\psi \rangle\langle \psi|i \rangle\Big|^2\notag\\
&=\frac{1}{d_{\textrm{A}}^2}~~\forall~~\psi.
\end{align}
We have completed the proof that a lower bound of $\mathcal{P}(\psi)\mathcal{F}(\psi)$ for an arbitrary initial state $|\psi\rangle$ is $d_{\textrm{A}}^{-2}$. \hfill $\blacksquare$

\section{Directly observing QIS from the average success probability of the PCTC}\label{sec:QIS_PCTC_proof}
Here, we show that the average success probability of the PCTC in our protocol is equal to the average value of the OTOC ($\mathcal{O}_\textrm{avg}$). Note that a similar proof can also be found in Refs.~\cite{Roberts2017,yoshida2019}. We begin by substituting Eq.~(\ref{seq:final_ket_Y_K}) into Eq.~(\ref{seq:P}) and averaging over all possible input states $\psi$, namely,
\begin{align}\label{seq:P_1}
\int d\psi \mathcal{P}(\psi) =& \int d\psi \langle f(\psi)|f(\psi)\rangle \notag\\
=& \int d\psi\times{}_{\textrm{A}}{\langle\psi|}\otimes{}_{\textrm{BB'}}{\langle\textrm{EPR}|}\otimes{}_{\textrm{HH'}}{\langle\textrm{EPR}|} \notag\\
&~~\times U^\textrm{T}_{\textrm{B'H'}}U^\dagger_{\textrm{AH}}(|\textrm{EPR}\rangle\langle\textrm{EPR}|_{\textrm{AB'}})U_{\textrm{AH}}U^*_{\textrm{B'H'}} \notag\\
&~~\times|\psi\rangle_\textrm{A}\otimes|\textrm{EPR}\rangle_\textrm{BB'}\otimes|\textrm{EPR}\rangle_\textrm{HH'}.
\end{align}
Note that the above EPR projection operator $|\textrm{EPR} \rangle\langle \textrm{EPR}|_\textrm{AB'}$ can be represented as a Haar average over all local unitary operators $W$ acting on the subsystems A and B'~\cite{Roberts2017,yoshida2019}:
\begin{equation}\label{seq:Haar}
|\textrm{EPR} \rangle\langle \textrm{EPR}|_\textrm{AB'} = \int dW W_\textrm{A} \otimes W^*_\textrm{B'}.
\end{equation}
Next, by substituting Eq.~(\ref{seq:Haar}) into Eq.~(\ref{seq:P_1}), one obtains
\begin{widetext}
\begin{align}\label{seq:P_2}
\int d\psi \mathcal{P}(\psi) &= \iint dW d\psi~{}_{\textrm{A}}{\langle\psi|}\otimes{}_{\textrm{BB'}}{\langle\textrm{EPR}|}\otimes{}_{\textrm{HH'}}{\langle\textrm{EPR}|}U^\textrm{T}_{\textrm{B'H'}}U^\dagger_{\textrm{AH}}W_{\textrm{A}}W^*_{\textrm{B'}}U_{\textrm{AH}}U^*_{\textrm{B'H'}} |\psi\rangle_\textrm{A}\otimes|\textrm{EPR}\rangle_\textrm{BB'}\otimes|\textrm{EPR}\rangle_\textrm{HH'} \notag\\
&=\iint dW d\psi~{}_{\textrm{A}}{\langle\psi|}\otimes{}_{\textrm{BB'}}{\langle\textrm{EPR}|}\otimes{}_{\textrm{HH'}}{\langle\textrm{EPR}|}(U^\textrm{T}_{\textrm{B'H'}}W^*_{\textrm{B'}}U^*_{\textrm{B'H'}})(U^\dagger_{\textrm{AH}}W_{\textrm{A}}U_{\textrm{AH}}) |\psi\rangle_\textrm{A}\otimes|\textrm{EPR}\rangle_\textrm{BB'}\otimes|\textrm{EPR}\rangle_\textrm{HH'} \notag\\
&=\iint dW d\psi~{}_{\textrm{A}}{\langle\psi|}\otimes{}_{\textrm{BB'}}{\langle\textrm{EPR}|}\otimes{}_{\textrm{HH'}}{\langle\textrm{EPR}|}(U^\dagger_{\textrm{B'H'}}W^\dagger_{\textrm{B'}}U_{\textrm{B'H'}})^\textrm{T}(U^\dagger_{\textrm{AH}}W_{\textrm{A}}U_{\textrm{AH}}) |\psi\rangle_\textrm{A}\otimes|\textrm{EPR}\rangle_\textrm{BB'}\otimes|\textrm{EPR}\rangle_\textrm{HH'} \notag\\
&=\iint dW d\psi~{}_{\textrm{A}}{\langle\psi|}\otimes{}_{\textrm{BB'}}{\langle\textrm{EPR}|}\otimes{}_{\textrm{HH'}}{\langle\textrm{EPR}|}(U^\dagger_{\textrm{BH}}W^\dagger_{\textrm{B}}U_{\textrm{BH}})(U^\dagger_{\textrm{AH}}W_{\textrm{A}}U_{\textrm{AH}}) |\psi\rangle_\textrm{A}\otimes|\textrm{EPR}\rangle_\textrm{BB'}\otimes|\textrm{EPR}\rangle_\textrm{HH'}.
\end{align}
\end{widetext}
Here, the last equality follows from Eq.~(\ref{seq:local_U_equality}). We can define the time-evolved version of $W$ in the Heisenberg picture according to the scrambling unitary operator $U$, i.e., $W_\textrm{A}(t) = U^\dagger_{\textrm{AH}}W_{\textrm{A}}U_{\textrm{AH}}$ and $W_\textrm{B}^\dagger(t) = U^\dagger_{\textrm{BH}}W^\dagger_{\textrm{B}}U_{\textrm{BH}}$. Furthermore, as we mentioned in Sec.~\ref{sec:QIS}, the input state $|\psi\rangle_\textrm{A}$ can be generated by a unitary operator $V_\textrm{A}$ acting on a fixed initial state $|\phi_0\rangle_\textrm{A}$, i.e., $|\psi\rangle_\textrm{A} = V_\textrm{A}|\phi_0\rangle_\textrm{A}$. Thus, the average over $\psi$ can be performed by integrating over $V$, namely,
\begin{align}\label{seq:P_3}
\int d\psi \mathcal{P}(\psi) =& \iint dW dV\times{}_{\textrm{A}}{\langle\phi_0|}\otimes{}_{\textrm{BB'}}{\langle\textrm{EPR}|}\otimes{}_{\textrm{HH'}}{\langle\textrm{EPR}|}\notag\\
&~~\times W_\textrm{B}^\dagger(t)V^\dagger_\textrm{A}W_\textrm{A}(t)V_\textrm{A}\notag\\
&~~\times |\phi_0\rangle_\textrm{A}\otimes|\textrm{EPR}\rangle_\textrm{BB'}\otimes|\textrm{EPR}\rangle_\textrm{HH'}\notag\\
=& \iint dW dV \langle W_\textrm{B}^\dagger(t)V^\dagger_\textrm{A}W_\textrm{A}(t)V_\textrm{A} \rangle.
\end{align}
As we mentioned in Sec.~\ref{sec:Protocol}, we relabel certain systems to simplify the previous proof. To better illustrate the connection between OTOCs and PCTCs through our decoding protocol, we return to the notation and labeling conventions employed in Fig.~\ref{fig:scheme}. Because the EPR projection operator is applied to the joint system EE' at $T_2$ [as shown in Fig.~\ref{fig:diagrammatic-proof}(a)], the unitary operator $W$ in Eq.~(\ref{seq:Haar}) should act on system E. Thus, Eq.~(\ref{seq:P_3}) can be rewritten as
\begin{equation}\label{seq:P_4}
\int d\psi \mathcal{P}(\psi) = \iint dW dV \langle W_\textrm{E}^\dagger(t)V^\dagger_\textrm{A}W_\textrm{E}(t)V_\textrm{A} \rangle=\mathcal{O}_\textrm{avg}.
\end{equation}
This completes the proof that the average value of the OTOC ($\mathcal{O}_\textrm{avg}$) is equal to the average success probability of the PCTC in our protocol. \hfill $\blacksquare$

The OTOC $\langle W^\dagger_\textrm{E}(t)V^\dagger_\textrm{A} W_\textrm{E}(t)V_\textrm{A} \rangle$ in Eq.~(\ref{seq:P_4}) can be interpreted as the overlap between the following two states:
\begin{align}
W_\textrm{E}(t)V_\textrm{A} & |\phi_0\rangle_\textrm{A}\otimes|\textrm{EPR}\rangle_\textrm{EE'}\otimes|\textrm{EPR}\rangle_\textrm{HH'}\notag\\
=& U^\dagger_{\textrm{HE}\rightarrow\textrm{BH}} W_\textrm{E} U_{\textrm{AH}\rightarrow\textrm{HE}} V_\textrm{A}\notag\\
&\times |\phi_0\rangle_\textrm{A}\otimes|\textrm{EPR}\rangle_\textrm{EE'}\otimes|\textrm{EPR}\rangle_\textrm{HH'},\label{seq:WV_state}\\
V_\textrm{A}W_\textrm{E}(t) & |\phi_0\rangle_\textrm{A}\otimes|\textrm{EPR}\rangle_\textrm{EE'}\otimes|\textrm{EPR}\rangle_\textrm{HH'}\notag\\
=& V_\textrm{A} U^\dagger_{\textrm{HE}\rightarrow\textrm{AH}} W_\textrm{E} U_{\textrm{BH}\rightarrow\textrm{HE}}\notag\\
&\times |\phi_0\rangle_\textrm{B}\otimes|\textrm{EPR}\rangle_\textrm{EE'}\otimes|\textrm{EPR}\rangle_\textrm{HH'}.\label{seq:VW_state}
\end{align}
The state in Eq.~(\ref{seq:WV_state}) follows the standard procedure of our protocol. Starting at $T_1$, Alice prepares the state $|\psi\rangle_\textrm{A}$ by applying the unitary operator $V_\textrm{A}$ on the initial state $|\phi_0\rangle_\textrm{A}$, and then the information is encoded in the joint system HE at time $T_2$ by the encoding scrambler $U_{\textrm{AH}\rightarrow\textrm{HE}}$. A unitary operator $W_\textrm{E}$ is then applied to system E, which can be regarded as a perturbation from the perspective of QIS. After system E travels backward in time from $T_2$ to $T_0$ through the PCTC, the decoding operation $U^\dagger_{\textrm{HE}\rightarrow\textrm{BH}}$ is applied, and Bob receives the state.

In contrast, the state in Eq.~(\ref{seq:VW_state}) represents the inverse version of our protocol, where the direction of all timelines (as indicated by the arrows in Fig.~\ref{fig:scheme}) is reversed. As time proceeds backward from $T_1$ to $T_0$, Bob encodes the state $|\phi_0\rangle_\textrm{B}$ into the joint system HE using the scrambling operation $(U^\dagger_{\textrm{BH}\rightarrow\textrm{HE}})^\dagger = U_{\textrm{BH}\rightarrow\textrm{HE}}$. After system E travels forward in time from $T_0$ to $T_2$ through the PCTC, the perturbation $W_\textrm{E}$ is applied to system E, and the temporal direction is again reversed at $T_2$. As time proceeds backward from $T_2$ to $T_1$, the decoding operation $U^\dagger_{\textrm{HE}\rightarrow\textrm{AH}}$ is applied, and Alice finally applies the unitary operator $V_{\textrm{A}}$ to her state.

Consequently, the value of the OTOC for the specific unitary operators $W_\textrm{E}$ and $V_\textrm{A}$ can be determined by calculating the overlap between the output states from the standard and inverse versions of our decoding protocol. Nevertheless, for the average value of the OTOC ($\mathcal{O}_\textrm{avg}$), we highlight that it can be directly obtained from the average success probability of the PCTC in our decoding protocol using Eq.~(\ref{seq:P_4}), without the need for the Haar averaging over all possible operators $W_\textrm{E}$ and $V_\textrm{A}$.

\section{Circuit decompositions of the three-qubit Clifford scrambling unitary operators for Quantinuum and IBM Quantum processors}\label{sec:Circuit_decomposition}
\begin{figure}
    \centering
    \includegraphics[width=\linewidth]{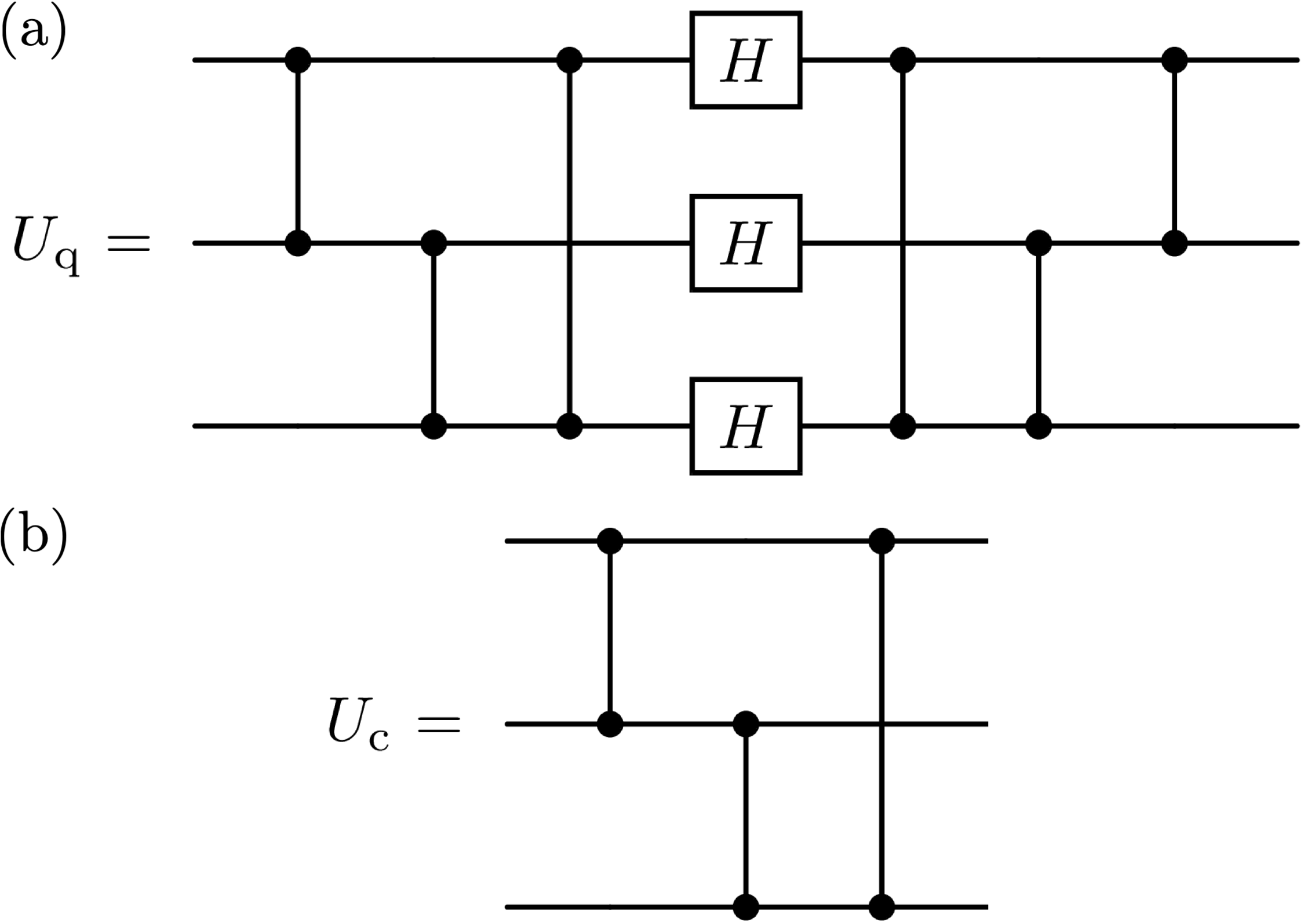}
    \caption{Circuit representation of two Clifford scramblers. (a) The quantum information scrambling unitary operator $U_\textrm{q}$. (b) The classical information scrambling unitary operator $U_\textrm{c}$.}
    \label{fig:Uq_Uc}
\end{figure}

In this Appendix, we provide the circuit decompositions for the three-qubit Clifford scrambling unitary operators ($U_\textrm{q}$ and $U_\textrm{c}$) used in the main text. In Ref.~\cite{yoshida2019}, the quantum information scrambler $U_\textrm{q}$ can be represented by several Hadamard gates, along with the controlled-$Z$ gates, as shown in Fig.~\ref{fig:Uq_Uc}(a). In contrast, the classical information scrambler $U_\textrm{c}$ only requires the three controlled-$Z$ gates, as shown in Fig.~\ref{fig:Uq_Uc}(b). 

We performed the experiment on the Quantinuum H1-1 quantum charge-coupled processor, which features 20 trapped-ion ($^{171}\textrm{Yb}^+$) qubits with all-to-all connectivity~\cite{Quantinuum}. This allows the circuit shown in Fig.~\ref{fig:Uq_Uc}(a) to be executed directly on the H1-1 processor. The calibration data for H1-1 is presented in Table~\ref{tab:Quantinuum-calibration}.

\begin{table}[!htpb]
    \centering
    \caption{Calibration data (obtained on 10 April 2024) for the Quantinuum processor H1-1 used in our experiments.}\label{tab:Quantinuum-calibration}
    \begin{tabular}{ll}
    \hline
    \hline
    Relaxation time ~&~ $\gg 1~\textrm{min}$\\
    Decoherence time ~&~ $\approx 4~\textrm{s}$\\
    Readout error ~&~ $2.5 \times 10^{-3}$\\
    Single-qubit gate error ~&~ $2.1 \times 10^{-5}$\\
    Two-qubit gate error ~&~ $8.8 \times 10^{-4}$\\
    \hline
    \hline
    \end{tabular}
\end{table}

\begin{figure*}
    \centering
    \includegraphics[width=\linewidth]{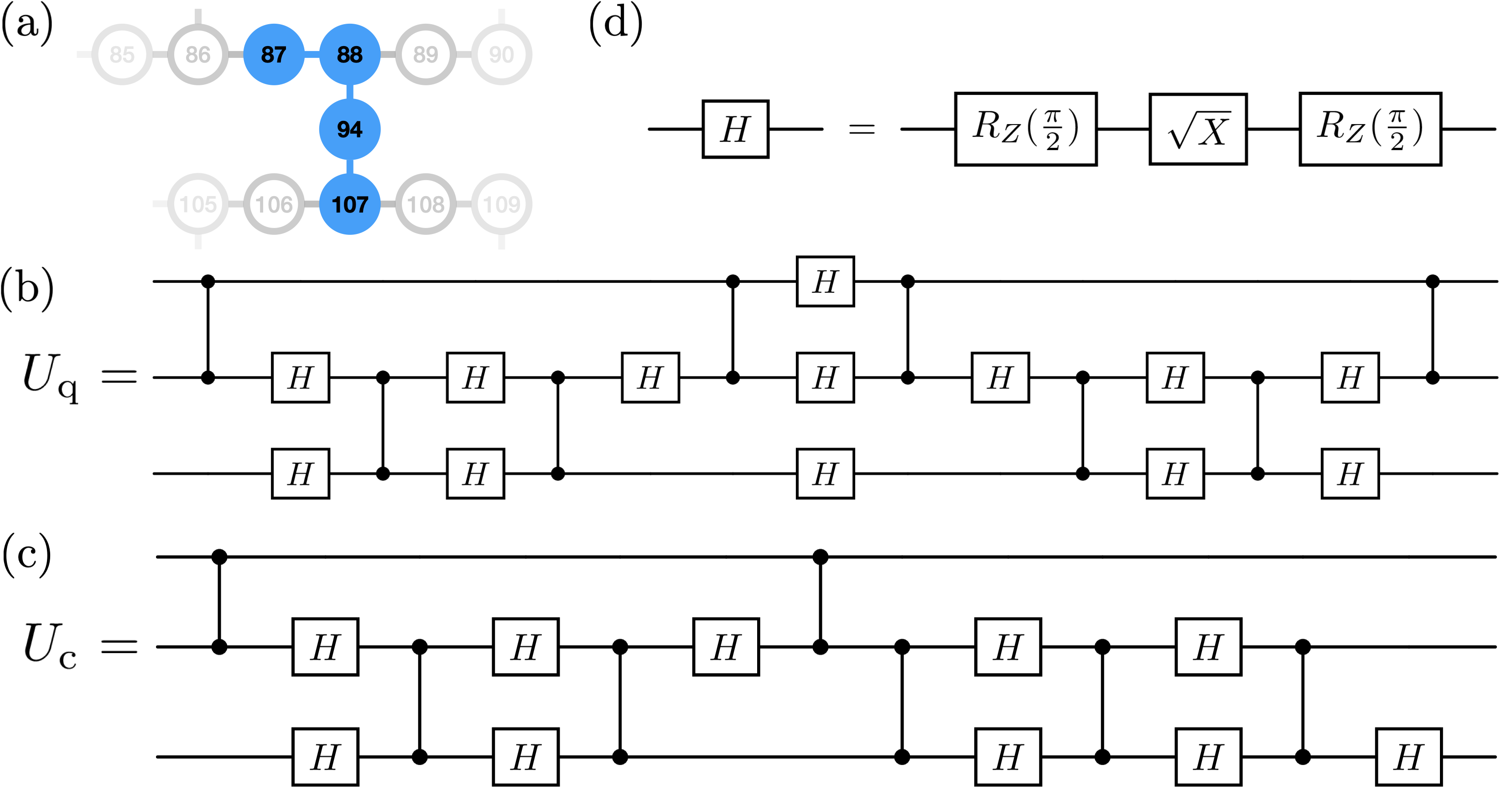}
    \caption{Topology of the IBM Quantum processor (ibm\_torino) and circuit decompositions used. (a) Topology of the ibm\_torino where the blue solid circles are the qubits used to conduct the experiment. (b) Circuit decomposition of the Clifford quantum information scrambler $U_{\textrm{q}}$ for ibm\_torino. (c) Circuit decomposition of the Clifford classical information scrambler $U_{\textrm{c}}$ for ibm\_torino. (d) The circuit decomposition of each Hadamard gate $H$ is given by a single $\sqrt{X}$ gate and two $R_Z$ (rotation around the $Z$-axis) gates with the rotational angle $\pi/2$.}
    \label{fig:ibm_decompose}
\end{figure*}

We also conduct the experiment on an IBM Quantum processor called ibm\_torino, which contains 133 superconducting qubits~\cite{IBMQ,Qiskit}. The topology of the qubits used in the experiment is illustrated in Fig.~\ref{fig:ibm_decompose}(a), and the corresponding calibration data for each used qubit is presented in Table~\ref{tab:IBM-calibration}. However, the two-qubit gates in Fig.~\ref{fig:Uq_Uc} do not match the topology shown in Fig.~\ref{fig:ibm_decompose}(a). Therefore, we provide circuit decompositions in Figs.~\ref{fig:ibm_decompose}(b) and \ref{fig:ibm_decompose}(c) to align with the topology of ibm\_torino, and the corresponding decomposition of each Hadamard gate is shown in Fig.~\ref{fig:ibm_decompose}(d).

\begin{table*}
    \centering
    \caption{Calibration data (obtained on 26 September 2024) for the IBM Quantum processor ibm\_torino used in our experiments. The qubit index refers to those in Fig.~\ref{fig:ibm_decompose}(a). Here, $R_Z$ is a single-qubit rotation gate around the $Z$ axis, and $\sqrt{X}$ is the square root of Pauli-$X$ gate. The average two-qubit (controlled-$Z$) gate error is estimated to be $1.62\times 10^{-3}$.}\label{tab:IBM-calibration}
    \begin{tabular}{ccccccccc}
    \hline\hline
    \multirow{2}{*}{\shortstack{System label\\(at time $T_0$)}} ~&~ \multirow{2}{*}{\shortstack{Qubit index}} ~&~ \multirow{2}{*}{\shortstack{Relaxation time\\($\mu$s)}} ~&~ \multirow{2}{*}{\shortstack{Decoherence time\\($\mu$s)}} ~&~ \multirow{2}{*}{\shortstack{Frequency\\(GHz)}} ~&~ \multirow{2}{*}{\shortstack{Anharmonicity\\(GHz)}} ~& \multicolumn{3}{c}{Errors} \\
    \cline{7-9}
     & & & & & & Readout & ~$R_Z$~ & $\sqrt{X}$\\
    \hline
    \multirow{2}{*}{H} & 87 & 293.88 & 256.22 & 0 & 0 & $6.30\times 10^{-3}$ & 0 & $1.44\times 10^{-4}$ \\
     & 88 & 280.95 & 403.23 & 0 & 0 & $2.54\times 10^{-2}$ & 0 & $1.49\times 10^{-4}$\\
    \hline
    E & 94 & 274.41 & 272.24 & 0 & 0 & $1.07\times 10^{-2}$ & 0 & $1.52\times 10^{-4}$\\
    \hline
    E' & 107 & 88.68 & 105.80 & 0 & 0 & $2.86\times 10^{-2}$ & 0 & $5.16\times 10^{-4}$\\
    \hline\hline
    \end{tabular}
\end{table*}

%

\end{document}